\documentclass[10pt, twocolumn, a4paper]{article}

\usepackage{amsmath,amssymb}
\usepackage[a4paper, top=2.0cm, bottom=1.5cm, left=1.5cm, right=1.5cm]{geometry}
\usepackage{helvet}
\usepackage{graphicx}
\usepackage{color}
\usepackage[colorlinks=true,bookmarks=false,citecolor=blue,urlcolor=blue]{hyperref}
\usepackage{textcomp}
\usepackage{physics}

\newcommand{\eps}{\varepsilon}      
\newcommand{\cra}[1]{\hat{a}^{\dag}_{#1}}  

\title{Topological photon pairs in a superconducting quantum metamaterial}

\author{Ilya~S.~Besedin$^{1,2}$, Maxim A. Gorlach$^{3}$, Nikolay~N.~Abramov$^{1,2}$, Ivan~Tsitsilin$^{1,2,4}$, Ilya~N.~Moskalenko$^{1,2}$, \\ Alina~A.~Dobronosova$^{5, 6}$, Dmitry~O.~Moskalev$^{5, 6}$, Alexey~R.~Matanin$^{5, 6}$, Nikita~S.~Smirnov$^{5, 6}$,\\ Ilya~A.~Rodionov$^{5, 6}$, Alexander N. Poddubny$^{3,7,8}$  and Alexey V. Ustinov$^{1,2,9}$}

\begin{document}

\small

\twocolumn[
\begin{@twocolumnfalse}
\maketitle

{\it \small{
$^1$~National University of Science and Technology MISIS, Moscow 119049, Russia\\
$^2$~Russian Quantum Center, Skolkovo, Moscow 143025, Russia\\
$^3$~Department of Physics and Engineering, ITMO University, Saint Petersburg 197101, Russia\\
$^4$~Moscow Institute of Physics and Technology, Dolgoprudny 141700, Russia\\
$^5$~FMN Laboratory, Bauman Moscow State Technical University, Moscow 105005, Russia\\
$^6$~Dukhov Automatics Research Institute (VNIIA), Moscow 127055, Russia\\
$^7$~Ioffe Institute, Saint Petersburg 194021, Russia\\
$^8$~Nonlinear Physics Centre, Australian National University, Canberra ACT 2601, Australia\\
$^9$~Physikalisches Institut, Karlsruhe Institute of Technology, Karlsruhe 76131, Germany
\\
}}
\end{@twocolumnfalse}]
\ \

\textbf{Recent discoveries in topological physics~\cite{Lu2014,Ozawa_RMP} hold a promise for disorder-robust quantum systems and technologies. Topological states provide the crucial ingredient of such systems featuring increased robustness to disorder and imperfections. Here, we use an array of superconducting qubits to engineer a one-dimensional topologically nontrivial quantum metamaterial~\cite{Macha2014}. By performing microwave spectroscopy of the fabricated array, we experimentally observe the spectrum of elementary excitations. We find not only the single-photon topological states but also the bands of exotic bound photon pairs~\cite{Mattis1986,Winkler} arising due to the inherent anharmonicity of qubits. Furthermore, we detect the  signatures of the two-photon bound edge-localized state which hints towards interaction-induced localization in our system. Our work demonstrates an experimental implementation of the topological model with attractive photon-photon interaction in a quantum metamaterial.}

Superconducting qubits are a viable platform for scalable quantum computers. Realization of quantum computation protocols and eventually quantum supremacy~\cite{Arute2019} relies largely on the coherent operation of ensembles of coupled qubits. A profound challenge in this direction is imposed by unavoidable parameter spread between the fabricated qubits of the ensemble. The need to control individual qubit parameters to reduce this spread makes scaling quantum computers up difficult. 

Even less practical appears the need of using individual qubit control tools for superconducting quantum metamaterials~\cite{Jung2014}~-- large arrays of interacting qubits, which can be viewed as an artificially created quantum matter. A promising way to tackle unavoidable irregularities in quantum metamaterials is provided by the concept of {\it topological states} whose existence is guaranteed by the global symmetries of the structure being insensitive to local imperfections~\cite{Lu2014,Ozawa_RMP}. In particular, recent studies suggest that a pair of photons propagating via a topological mode preserves quantum correlations significantly longer than a photon pair propagating in the bulk~\cite{Blanco-Science,Wang2019,Blanco-nanoph}. 

Photon-photon interactions arising due to the nonlinearity of the medium provide an exciting additional degree of freedom giving rise to a plethora of topological phenomena including interaction-induced topological states. One of the simplest  interacting models is the paradigmatic Bose-Hubbard model which arises in various contexts including cold atom ensembles in optical lattices~\cite{Jaksch2005}, magnetic insulators~\cite{Giamarchi2008} as well as arrays of transmon qubits. The inherent anharmonicity of transmon qubit potential enables  strong  photon-photon interactions~\cite{Carusotto2020}. Combining this feature with fine-tuning of qubit eigenfrequency or temporal modulation of qubit couplings, one can switch the system into the quantum Hall phase~\cite{Roushan:2016NatPhys,Roushan-Science}. Note that the elementary bosonic excitations in qubit array are, strictly speaking, plasmon-polaritons as they are superpositions of plasma oscillations in Josephson junctions and distributed electric fields in their environment~\cite{Hartmann2010} and we term them ``photons'' only for the sake of simplicity. 

An interesting feature of the Bose-Hubbard model is the emergence of bound boson pairs (doublons) mediated by the interactions~\cite{Mattis1986,Winkler}. Such quasiparticles can localize at the edge of the system forming a two-photon bound edge state which arises even in the absence of single-photon localized states~\cite{Gorlach-H-2017,Olekhno}. The topological physics of doublons has been extensively explored in a series of recent theoretical works~\cite{Gorlach-2017,Salerno,Salerno2020,Stepanenko2020} and also emulated with a classical system of higher dimensionality~\cite{Olekhno}. However, the experimental investigation of topological properties of bound photon pairs is still lacking.

\begin{figure*}[t]
    \centerline{\includegraphics[width=16cm]{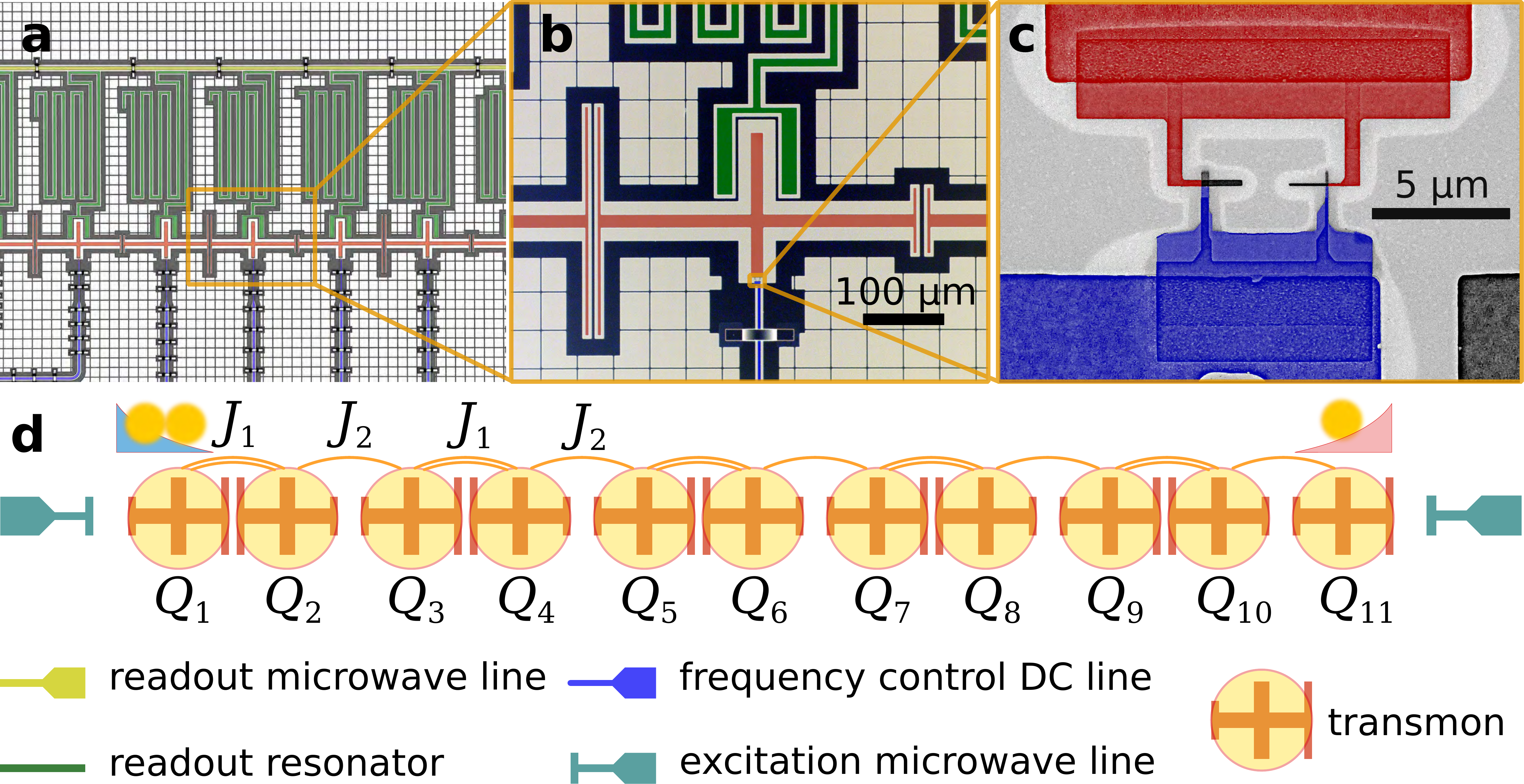}}
    \caption{\footnotesize {\bf Implementation of the Su-Schrieffer-Heeger model with the array of transmon qubits.} {\bf a}, False color image of the array of transmon qubits. {\bf b}, Enlarged fragment showing the geometry of the individual transmon qubit. {\bf c}, False-color SEM image of transmon junctions. {\bf d}, The sketch of the designed structure showing the effective coupling amplitudes and the location of single-photon topological state and two-photon bound edge-localized state.}
    \label{fig:System}
\end{figure*}

To fill this gap and to highlight the important aspects of topological protection in interacting quantum models,  we design and investigate experimentally a dimerized array of qubits with alternating nearest-neighbor coupling strengths [Fig.~\ref{fig:System}(a-c)]. This quantum metamaterial, depicted schematically in Fig.~\ref{fig:System}(d) provides a realization of the well-celebrated topologically nontrivial Su-Schrieffer-Heeger model (SSH)~\cite{Heeger1988} which has been implemented in a variety of systems ranging from simple mechanical~\cite{Kane2014} or microwave~\cite{Slob} structures to photonic~\cite{Malkova:09}, polaritonic~\cite{StJean} and cold atom~\cite{Leseleuc2019} realizations. Importantly, in contrast to the previous studies of one-dimensional topological qubit arrays~\cite{Cai2019}, we probe here not only the single-photon excitations but also the two-photon ones, examining such exotic phenomena as repulsively bound photon pairs and their interaction-induced localization. 

The circuit that we studied consists of $N=11$ nearest-neighbor coupled superconducting transmon qubits described by the Bose-Hubbard Hamiltonian
\begin{gather}
\mathcal{\hat H}_{\rm{BH}}/h=\sum\limits_{q=1}^{N}\,\left[f_q\,\hat{n}_q +\frac{\delta_q}{2}\,\hat{n}_q \,(\hat{n}_q-1)\right]
 \notag\\
+\sum\limits_{q=1}^{(N-1)/2}\,\left[J_1\,\hat a_{2q}\,\hat a_{2q-1}^\dag+J_2\,\hat a_{2q}\,\hat a_{2q+1}^\dag\right]+\text{H.c.}\:,\label{BoseH}
\end{gather}
where $\hat{n}_q=\hat a^\dag_q \hat{a}_q$ is photon number operator, $f_q$ is the eigenfrequency of $q^{\rm{th}}$ transmon which can be flexibly tuned in the range from $3.73$ to $3.82$~GHz, $\delta_q=-155$~MHz is qubit anharmonicity responsible for the effective photon-photon interaction, and the amplitudes $J_1=55.1$~MHz and $J_2=17.1$~MHz describe the nearest-neighbor coupling of qubits. All qubits are tuned to the same frequency equal to $f_0=3.8$~GHz and $f_0'=3.75$~GHz in the single-photon and two-photon experiments, respectively. To probe the modes of qubit array, we apply a monochromatic driving  with frequency $f_d$ and amplitude $\Omega$. In our experiments, the driving is applied to one of the edges of qubit array, either to $s=1$ or to $s=11$, while the number of photons can be measured in any qubit.

In the absence of driving and dissipation, the Bose-Hubbard model conserves the number of photons. Therefore, depending on the driving amplitude, one can address either single-photon sector of the Hilbert space, or the two-photon eigenstates, which we exploit to investigate few-boson physics in the interacting Bose-Hubbard model.



\begin{figure*}[t]
    \centerline{\includegraphics[width=18cm]{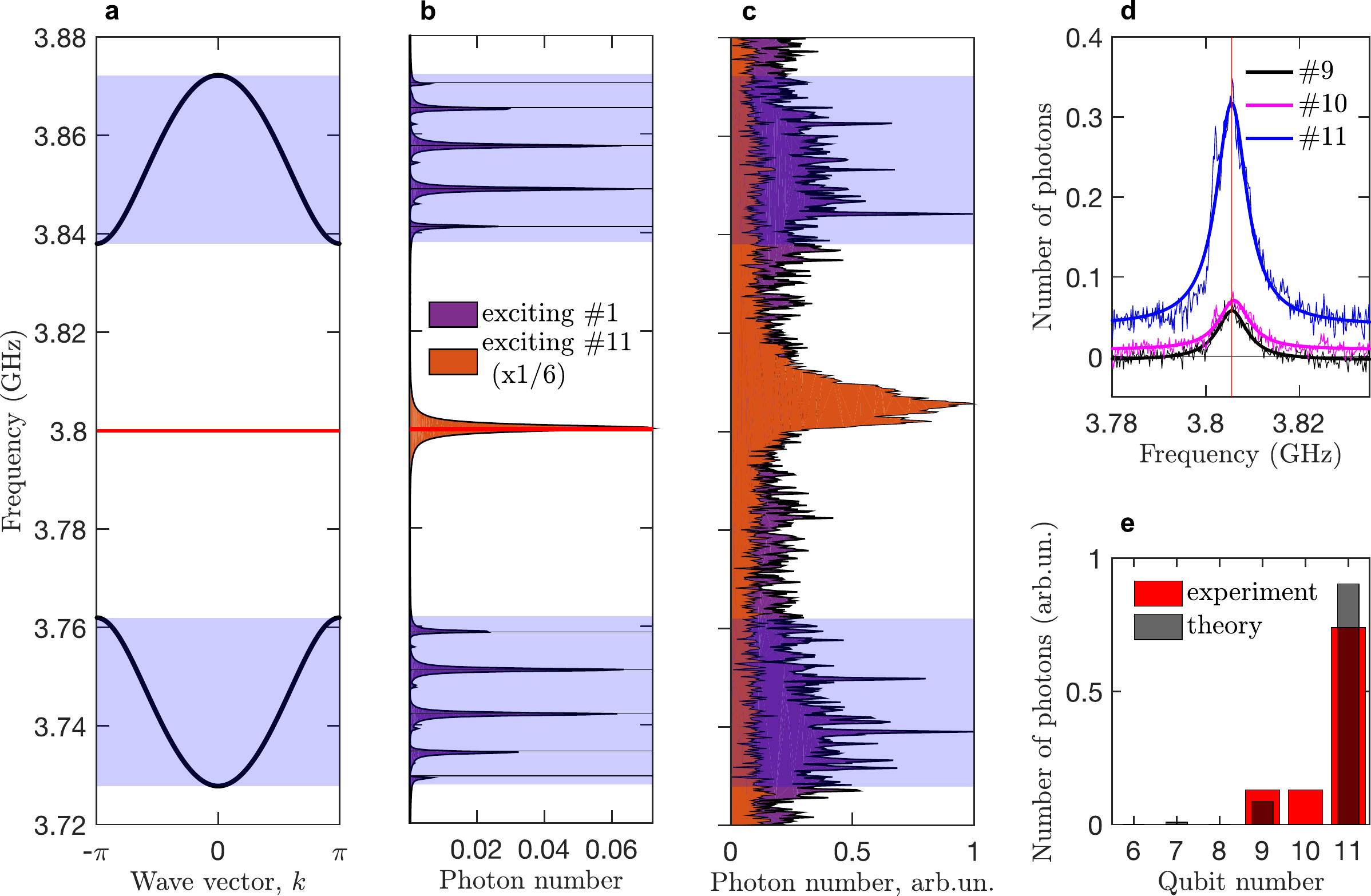}}
    \caption{\footnotesize {\bf Spectroscopy of single-photon excitations in qubit array.} {\bf a}, Dispersion of single-photon excitations in the array of qubits tuned to the same frequency $f_0=3.8$~GHz. Shaded areas illustrate the boundaries of the bulk bands. Horizontal line at the frequency $3.8$~GHz depicts the spectral position of the single-photon topological state. {\bf b}, Calculated spectra showing the average number of photons in the array versus driving frequency for excitation of the first (magenta color) or last (orange color) qubits of the array. Calculation details are provided in Sec.~II of Supplementary Materials. {\bf c}, Experimental spectra showing the dependence of normalized average photon number in the array versus driving frequency. Magenta and orange colors correspond to the excitation of the first and the last qubits in the array, respectively. {\bf d}, Average photon number measured in qubits No. 9, 10 and 11 versus driving frequency upon excitation of qubit No. 11. {\bf e}, Single-photon probability distribution for the edge state extracted from the amplitudes of the peaks measured for qubits No.~9, 10 and 11 and depicted in panel {\bf d}. }
    \label{fig:SinglePhoton}
\end{figure*}

For the single-photon excitations, the interaction term $\propto\delta_q$ in the Hamiltonian Eq.~\eqref{BoseH} is not manifested, and the physics of this system is captured by the Su-Schrieffer-Heeger model. Therefore, we expect two bands of single-photon states with the dispersion given by
\begin{equation}\label{BulkSinglePhoton}
f=f_0\pm\sqrt{J_1^2+J_2^2+2\,J_1\,J_2\,\cos k}\:,
\end{equation}
as illustrated in Fig.~\ref{fig:SinglePhoton}(a). Here, the Bloch wave number $k$ ranges from $-\pi$ to $\pi$ taking quantized values in a finite system. As a result, exciting the system at the first qubit and measuring the expectation value of the photon number operator $\left<\hat n_1\right>$, we expect a series of distinct peaks corresponding to the bulk modes of the array [Fig.~\ref{fig:SinglePhoton}(b)] emergent in the experimental data [Fig.~\ref{fig:SinglePhoton}(c)].

Single-photon excitations described by Eq.~\eqref{BulkSinglePhoton} are spread over the entire array. However, besides such delocalized excitations the system also supports an edge-localized topological state with the frequency precisely in the middle of bandgap, $f_{\rm{edge}}=f_0$. The wave function of this state exhibits  an exponential localization at the edge of the array with the weak tunneling link, i.e. at 11$^{\rm{th}}$ qubit. Therefore, exciting the array at the last qubit, we couple predominantly to the edge state manifested as a sharp peak in the middle of the bulk bandgap [Fig.~\ref{fig:SinglePhoton}(b), orange color]. A similar peak is observed in experimental data [Fig.~\ref{fig:SinglePhoton}(c)] providing direct evidence of the single-photon topological state in the SSH-type array of coupled qubits.

Besides providing the spectroscopic evidence of a single-photon topological state, we also examine the probability distribution for the observed mode. To this end, we excite the array at the 11$^{\rm{th}}$ qubit and measure the expectation value of photon number operator in qubits No.~11, 10 and 9 versus driving frequency, as shown in Fig.~\ref{fig:SinglePhoton}(d). All three curves have a similar shape featuring a pronounced peak at the same frequency corresponding to the resonant excitation of the single-photon edge state. The amplitudes of these peaks are related to the eigenmode profile, which enables us to extract the probability distribution for the edge mode directly from the experimental data. The extracted results are compared with predictions of the tight-binding model in the histogram Fig.~\ref{fig:SinglePhoton}(e). As expected, the probability amplitude decays away from the edge of the array, which clearly confirms the edge-localized nature of the observed single-photon state.


The two-photon modes of the designed system feature more complex behavior than their single-photon counterparts and belong to four major types~\cite{Gorlach-2017,DiLiberto}. The majority of the two-photon eigenstates is presented by the scattering states with the energy given by the sum of single-photon energies in the continuum limit. The second group is the single-photon edge states, where one photon is localized at the weak link edge of the array, whereas the second photon propagates along the array. These two types of modes strongly resemble those in the single-particle SSH model.

\begin{figure*}[t]
    \centerline{\includegraphics[width=18cm]{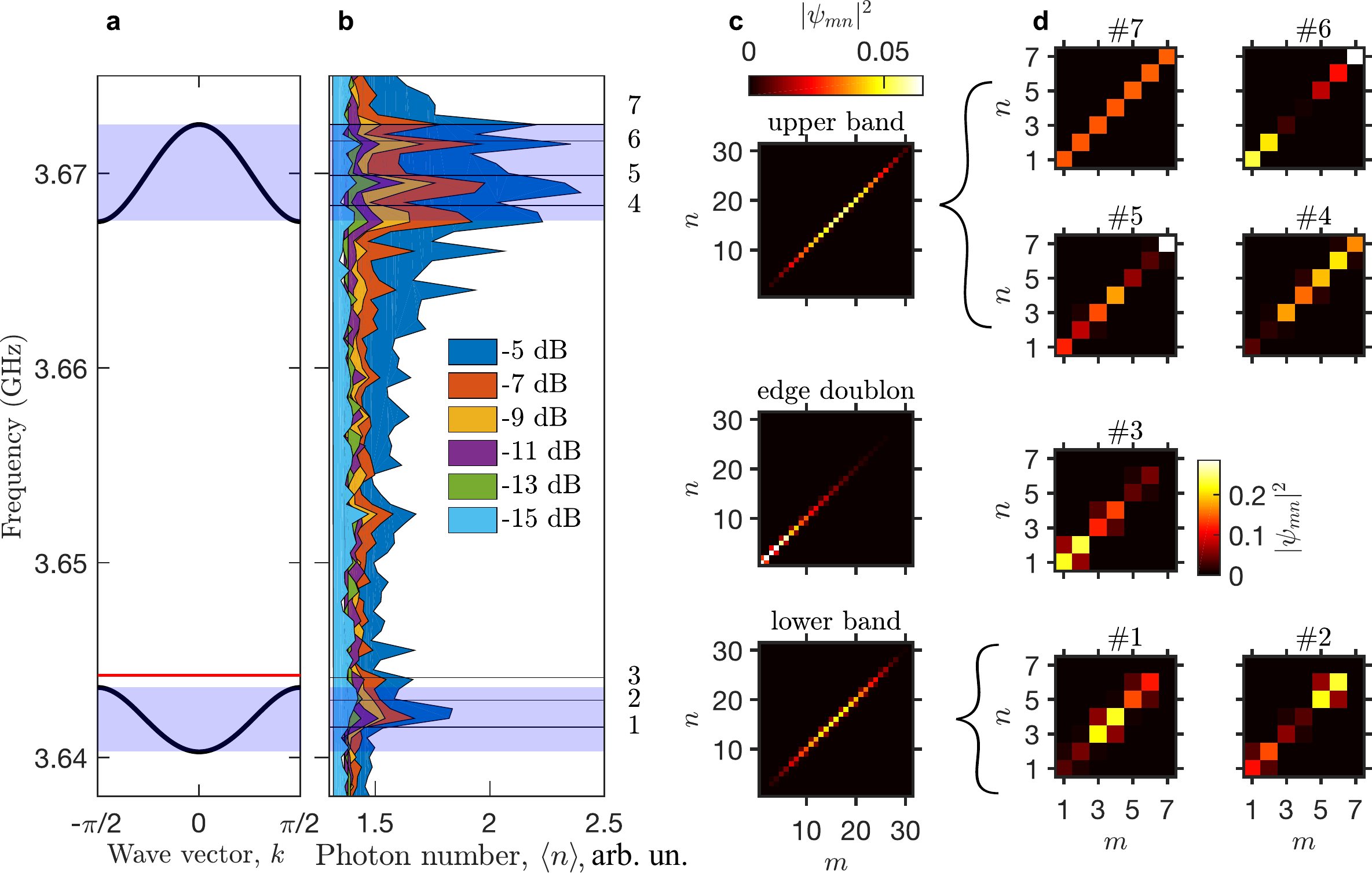}}
    \caption{\footnotesize {\bf Spectroscopy of two-photon excitations in qubit array.} {\bf a}, Calculated  dispersion of bound photon pairs. Higher-frequency two-photon scattering states are not shown. Red horizontal line shows the spectral position of the doublon edge state localized at the opposite edge of the array compared to the single-photon state. {\bf b}, Experimental spectrum of normalized average photon number in the array when the first qubit is pumped. Shaded areas show the boundaries of bulk doublon bands. In this experiment, the array consists of $7$ qubits tuned to resonance at $3.75$~GHz. {\bf c}, Calculated two-photon probability distribution in the array of 31 qubit for the three representative cases: upper bulk doublon band (top), doublon edge state (middle), lower bulk doublon band (bottom). {\bf d}, Calculated two-photon probability distributions for the array of 7 qubits. Numbers near each of panels correspond to the characteristic peaks marked in panel {\bf b}. Panel 3 shows the signatures of interaction-induced edge localization.}
    \label{fig:TwoPhoton}
\end{figure*}

However, the effects of interaction give rise to two novel types of modes associated with bound photon pairs (doublons) as well as their edge states. The bulk spectrum of the system features four doublon bands~\cite{Gorlach-2017,DiLiberto}. Two doublon bands have the frequency which scales linearly with $\delta$ being detuned from the scattering continuum approximately by $\delta=-155$~MHz. The dispersion of these two doublon bands calculated by the modified Bethe ansatz method~\cite{Gorlach-2017} is presented in Fig.~\ref{fig:TwoPhoton}(a). While bulk doublon dispersion strongly resembles that of the SSH model, the doublon edge state obtained in our calculations [Fig.~\ref{fig:TwoPhoton}(a)] exhibits quite unexpected features. Contrary to the single-photon scenario, it is localized at the strong link edge of the array providing an example of interaction-induced localization.

Due to the good spectral isolation of the discussed doublon bands, the respective two-photon states can be probed spectroscopically. For that purpose, we apply the driving signal to the first qubit of the array measuring the spectrum of photon number amplitudes  $\left<\hat n_1\right>$ [Fig.~\ref{fig:TwoPhoton}(b)]. While at low input powers we do not observe any pronounced features in the spectrum, the situation changes dramatically when input power is increased up to the level allowing two-photon transitions in qubits. In agreement with the calculated positions of bulk doublon bands, two separate groups of peaks emerge.

To provide a clearer interpretation of the observed peaks, we perform the numerical diagonalization of the Hamiltonian for a larger system composed of $31$ qubit [Fig.~\ref{fig:TwoPhoton}(c)]. The upper group of peaks is related to ``antisymmetric'' doublon states when the phase of the doublon wave function for the adjacent positions of bound photon pair differs by $\pi$, whereas the lower group is associated with ``symmetric'' doublon states, when the respective phase difference is equal to $0$. 

The doublon edge state emerges at the upper boundary of the lower band and features relatively  weak interaction-induced localization with the localization length exceeding 10 sites. Surprisingly, even though this  localization
fully shows up only in the long arrays with $N\gtrsim 20$ qubits, the corresponding state is predominantly  localized at the edge already for  odd numbers of qubits $N\gtrsim 5$ [see Supplementary Materials, Sec.~V]. The reason for the persistence of localization for short arrays with odd $N$ is that the doublons localize only at the strong-link edge, and thus the hybridization of the states from the opposite edges does not mask the localization.

Comparing our theoretical predictions with the experimental results  for the  array of $7$ resonant qubits [Fig.~\ref{fig:TwoPhoton}(d)], we identify the states \#1-2 associated with the lower doublon band and states \#4-7 related to the upper doublon band. Finally, the state \#3 heralds the emergence of doublon edge states.
The obtained results are thus pointing towards interaction-induced localization in our system.

To conclude, our experiments demonstrate the power of qubit arrays for exploring few-photon topological physics. As we prove, the two-photon spectrum is dramatically enriched  compared to the non-interacting scenario. Interactions give rise to exotic bound photon pairs whose energies and wave functions can be probed spectroscopically. Furthermore, our system exhibits the signatures of interaction-induced two-photon localization. This highlights a promise of interacting models for quantum topological photonics paving a way towards disorder-robustness of quantum metamaterials.

\section*{\small Methods}

{\scriptsize

\textbf{Tight binding simulations.}
To find the eigenstates of the Bose-Hubbard Hamiltonian Eq.~\eqref{BoseH}, we use the fact that the Hamiltonian $\mathcal{\hat H}_{\rm{BH}}$ commutes with the operator $\hat{n}=\sum_q\,\hat{n}_q$. This means that the total number of photons in this model is conserved, and the two-photon wave function can be presented in the form
\begin{equation}\label{TwoPhotonFunction}
\ket{\psi}=\frac{1}{\sqrt{2}}\,\sum\limits_{m,n}\,\beta_{mn}\,\cra{m}\,\cra{n}\,\ket{0}\:,
\end{equation}
where $\beta_{mn}=\beta_{nm}$ are the unknown superposition coefficients. The two-photon eigenstates are found from the Schr\"odinger equation $\mathcal{\hat H}_{\rm{BH}}\,\ket{\psi}=\eps\,\ket{\psi}$. Combining Eqs.~\eqref{BoseH} and \eqref{TwoPhotonFunction}, we obtain the linear system of equations for the unknown superposition coefficients:
\begin{gather}
\left(\eps-2U\right)\,\beta_{nn}=-2J_{n-1}\,\beta_{n-1,n}-2J_n\,\beta_{n,n+1}\:,\\
\eps\,\beta_{mn}=-J_{m-1}\,\beta_{m-1,n}-J_{n-1}\,\beta_{m,n-1}\notag\\
-J_m\,\beta_{m+1,n}-J_n\,\beta_{m,n+1}\:,
\end{gather}
where $m<n$, $\beta_{0n}=0$, $J_n=J_1$ for odd $n$ and $J_n=J_2$ for even $n$. Solving this system numerically, we find the two-photon eigenstates supported by the finite array. The frequencies shown in Figs.~\ref{fig:SinglePhoton}, \ref{fig:TwoPhoton} correspond to $\eps/(2\,h)$.

\textbf{Sample fabrication.}

The sample is made in a three-stage process: (I) base Al layer patterning, (II) Josephson junction double-angle evaporation and lift-off, (III) low impedance crossover fabrication to suppress stray microwave chip modes. We use a high-resistivity intrinsic silicon substrate ($\rho > 10000 \mathrm{\Omega}\cdot$~cm, 500~$\mu$m thick) prepared by Piranha-based wet cleaning and HF dip to remove surface oxide damage. The qubit capacitors, ground plane, readout resonators, and control wiring are made using e-beam evaporated 100-nm-thick epitaxial Al \cite{Rodionov2019} and subsequent patterning by means of direct laser lithography and $\mathrm{B}\mathrm{Cl}_3$/$\mathrm{Cl}_2$ inductively coupled plasma etch. The native Al oxide is removed with in-situ Ar-ion milling, followed by Al-AlOx-Al Josephson junction e-beam shadow evaporation (±11º, 25/45 nm) and lift-off in N-methyl-2-pyrrolidone at 80\textdegree{}C. Finally, low impedance free-standing crossovers are fabricated in a process similar to the fabrication steps outlined in Ref. \cite{Chen2014}, with an important improvement: we have used Al dry etching ($\mathrm{Cl}_2$-based) to pattern crossovers instead of wet process and redesign crossover topology to eliminate the Al base layer damage and provide mA-range crossover critical current.

\textbf{Measurement scheme.}
We perform continuous wave two-tone spectroscopy by measuring the shift of the scattering amplitude at the frequency of the unperturbed readout resonator $\Delta S_{21}(f_r)$, while applying a monochromatic drive tone with amplitude $\Omega$ and frequency $f_d$. The drive tone is applied to one of the edges of the array.  The readout circuitry of the sample is designed such that the scattering amplitude shift is proportional to the energy stored in the corresponding transmon. 

The first condition for this regime of operation is that all dispersive shifts $\chi$ are at least order of magnitude less than the resonator linewidths, yielding the first-order expansion
\begin{equation}
\label{eq:DeltaS21}
\langle \Delta S_{21}(f_r)\rangle = \frac{\partial S_{21}^{\mathrm{notch}}(f, f_r, Q_l, Q_c)}{\partial f_r} \chi_q \langle n \rangle,
\end{equation}
where $S_{21}^{\mathrm{notch}}(f, f_r, Q_l, Q_c)$ is the theoretical frequency dependence of the $S_{21}$ parameter for resonator frequency $f_r$, its loaded and external quality factors $Q_l$ and $Q_c$, respectively~\cite{Probst2015}, and $\langle n \rangle$ is the mean number of photons in transmon. 

The second condition requires that the anharmonicity of the transmons is order of magnitude less than the qubit-transmon detuning \[ |\delta_q| \ll |f_r-f_q|\:. \] To the first order in $\delta$ and second order in the qubit-resonator coupling constant $g$, the dispersive shift is given by
\begin{equation}
\label{eq:chi}
\chi_q = -\frac{2g^2\delta_q}{\left(f_r-f_q\right)^2}.
\end{equation}

For the single-photon state measurements, we convert the scattering amplitude shift into mean photon number, computing the coefficient from Eqs.~\eqref{eq:DeltaS21}, \eqref{eq:chi}.

The bare frequencies of the qubits are set to their target values using individual DC flux control lines, each coupled to one of the qubits. Mutual inductances between the flux lines and SQUIDs of the corresponding qubit, as well as its nearest neighbors, frozen-in flux values, precise values of Josephson junction critical currents, and qubit-resonator coupling have been extracted from fitting a series of two-tone calibration measurements to a linear oscillator model (see Supplementary Materials, Sec.~IV). The values of $J_1, J_2$ and $\delta$ are determined from electrostatic simulation. 

}

\subsection*{\small Data availability}
The data that support the findings of this study are available from the corresponding author upon request.

\subsection*{\small Acknowledgments}
Development of theoretical models was supported by the Russian Foundation for Basic Research (grant No.~18-29-20037). The Russian Science Foundation supported the experiments (contract No. 16-12-00095) and numerical simulations (grant No.~16-19-10538). A.N.P. and M.A.G. acknowledge partial support by the Foundation for the Advancement of Theoretical Physics and Mathematics ``Basis''. A.V.U. acknowledges partial support by the Deutsche Forschungsgemeinschaft (DFG) by the Grant No. US 18/15-1. Sample studied in this work was made at the BMSTU Nanofabrication Facility (FMN Laboratory, FMNS REC, ID 74300). MISIS team also acknowledges support from the Ministry of Education and Science of the Russian Federation in the framework of the Increase Competitiveness Program of the National University of Science and Technology MISIS (contract No. K2-2020-017).

\subsection*{\small Author contributions}
M.A.G. and A.N.P. elaborated the theoretical models. I.S.B., A.N.P. and M.A.G. performed numerical simulations. A.A.D., D.O.M., A.R.M., N.S.S. and I.A.R. fabricated the sample. I.T. and I.S.B. designed and simulated the chip. N.A. built the multi-channel frequency control hardware. I.S.B., I.T., and I.N.M. performed the measurements. A.V.U. supervised the project.

\subsection*{\small Competing interests}
The authors declare that they have no competing interests.

\subsection*{\small Additional information}
Correspondence and requests for materials should be addressed to M.A.G. (email: m.gorlach@metalab.ifmo.ru).

\bibliographystyle{naturemag}
{\scriptsize
\bibliography{TopologicalLib2}
}

\end{document}


\maketitle

\begin{affiliations}
\item National University of Science and Technology MISIS, Moscow 119049, Russia\\
\item Russian Quantum Center, Skolkovo, Moscow 143025, Russia\\
\item Department of Physics and Engineering, ITMO University, Saint Petersburg 197101, Russia\\
\item Moscow Institute of Physics and Technology, Dolgoprudny 141700, Russia\\
\item FMN Laboratory, Bauman Moscow State Technical University, Moscow 105005, Russia\\
\item Dukhov Automatics Research Institute (VNIIA), Moscow 127055, Russia\\
\item Ioffe Institute, Saint Petersburg 194021, Russia\\
\item Nonlinear Physics Centre, Australian National University, Canberra ACT 2601, Australia\\
\item Physikalisches Institut, Karlsruhe Institute of Technology, Karlsruhe 76131, Germany
\end{affiliations}

\tableofcontents

\newpage

\section{Formation of the doublon bands}\label{sec:DoublonBands}

\begin{figure*}[h]
	\centerline{\includegraphics[width=0.83\linewidth]{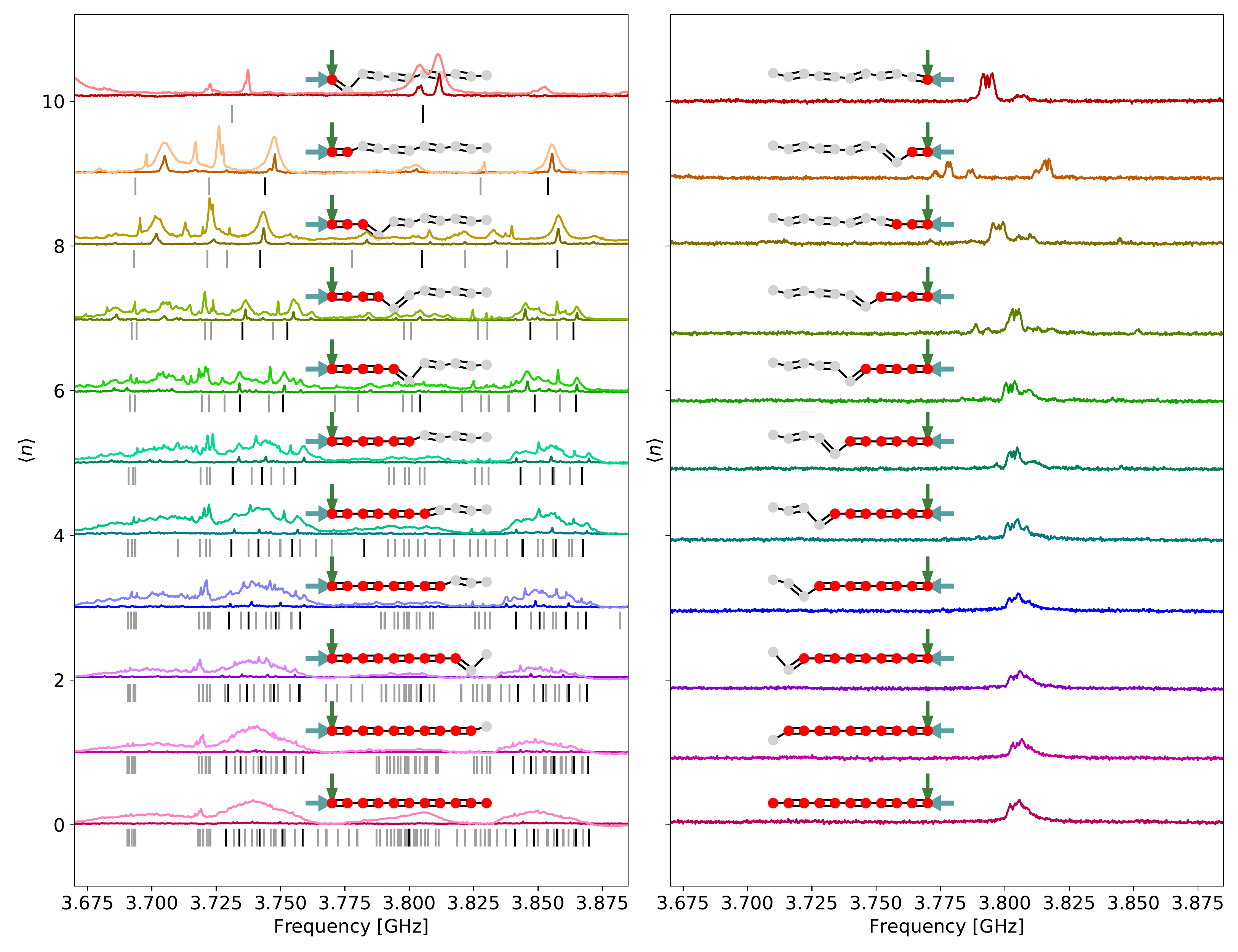}}
	\caption{\footnotesize {\bf Two-tone spectroscopy of bulk and edge states in the array of transmon qubits for the different effective lengths of the array}. Light curves correspond to strong drive (0 dB) and dark curves corrspond to weak drive (-20 dB). The pictogram above each curve shows the scheme of experimental setup, where red and grey circles depict qubits tuned into resonance and out of resonance, respectively. Vertical position of the circle shows whether the frequency of a given transmon is above or below the resonance frequency. Horizontal teal blue arrows indicate the transmon which is excited, and the vertical green arrows point at transmon which is measured. Black and grey vertical lines below each graph depict the calculated frequencies of single-photon and two-photon states.}
	\label{im:all-spectra}
\end{figure*}

Results of two-tone spectroscopy of the designed transmon array measured for the different effective lengths are presented in Fig.~\ref{im:all-spectra}. We perform two types of experiments: (a) excite the first qubit of the array at the strong link edge and measure the average photon number in it; (b) excite the last qubit of the array at the weak link edge and measure the average photon number in it. 

In the first case, we observe a complicated interplay of single-photon and two-photon modes. In the experiment, these two types of modes can be distinguished by their dependence on the driving power: while single-photon states are manifested already for relatively weak driving, the spectroscopy of the two-photon modes requires essentially higher driving power. Increasing the length of the array up to 11 qubits, we observe that the distinct peaks merge into groups highlighting the formation of the bulk bands. Three broad peaks shown in this panel correspond to the so-called scattering states\cite{Gorlach-2017} with the energy given by the sum of the single-photon energies in the continuum limit. The peaks corresponding to the bulk doublon bands are red-shifted with respect to the scattering states and discussed in more detail in the manuscript main text.

In contrast, in the second case corresponding to the excitation of the last qubit in the array, we observe a single dominant peak at frequencies around $3.8$~GHz which is the resonance frequency of the individual qubits of the array. This single peak is associated with the single-photon edge state localized at the weak link edge and having good spatial overlap with the pump.

Note that in our experiments each spectrum is recorded 10 times to mitigate the effect of slow frequency drifts of the resonance frequency of the readout resonators. The mean standard deviation of the average photon population $\langle n \rangle$ in all measurements is 0.023.


\section{Simulation of average transmon occupancy}\label{sec:SSHmap}

An array of driven transmons with tunable frequency where each transmon is coupled to its nearest neighbors via alternating mutual capacitances is described by the Hamiltonian
\begin{equation}
\label{eq:Hamiltonian}
\begin{split}
\mathcal{\hat H} /h = & \sum\limits_{q=1}^{N}\,\left[f_q\,\hat{n}_q +\frac{\delta_q}{2}\,\hat{n}_q \,(\hat{n}_q-1)\right]\\
& +\sum_{q=1}^{(N-1)/2}\,\left\lbrace \sqrt{f_{2q}}\,(\hat a_{2q} + \hat a_{2q}^\dag)\,\left[t_{1}\sqrt{f_{2q-1}}\,(\hat a_{2q-1} + \hat a_{2q-1}^\dag)\right.\right.
\left.\left.+t_{2}\sqrt{f_{2q+1}}\,(\hat a_{2q+1} + \hat a_{2q+1}^\dag) 
\right]\right\rbrace\\
& + \frac{\Omega}{2\pi}\cos(2\pi f_d t)(\hat a_{s} + \hat a_{s}^\dag)\:,
\end{split}
\end{equation}
where $s$ is the number of the driven transmon, $s=1$ or $s=11$. All transmon qubits are tuned to the same resonance frequency $f_q=f_0$ which results in alternating coupling constants $J_{1,2}=f_0\,t_{1,2}$. To obtain the steady-state solution, we first transform the driven Hamiltonian Eq.~\eqref{eq:Hamiltonian} by applying the unitary transformation 
\begin{equation}
\hat{U} = e^{2\pi i f_d t\sum\limits_{q}\hat{n}_q}
\end{equation}
and calculating the transformed Hamiltonian as
\begin{equation}
\mathcal{\hat H}'=\hat{U}\mathcal{\hat H}\,\hat{U}^{-1}-i\hbar\,\hat{U}\,\frac{\partial \hat{U}^{-1}}{\partial t}\:.
\end{equation}
After counter-rotating terms are dropped, this results in the time-independent Hamiltonian
\begin{equation}
\begin{split}
\mathcal{\hat H}_\mathrm{RWA} /h = \sum\limits_{q=1}^{N} & \left[ (f_q-f_d)\hat{n}_q +
\frac{\delta}{2}\hat{n}_q (\hat{n}_q-1) \right ]
+ \\
\sum_{q=1}^{(N-1)/2} & \left[J_1 (\hat{a}_{2q}^\dag\,\hat a_{2q-1} + \hat a_{2q-1}^\dag \hat{a}_{2q}) + J_2\,(\hat{a}_{2q}^\dag \hat a_{2q+1} + \hat a_{2q+1}^\dag\hat{a}_{2q})\right]  \\
& +\frac{\Omega}{4\pi}(\hat a_{s} + \hat a_{s}^\dag)\:.
\end{split}
\end{equation}

Decoherence and, specifically, decay, is a crucial ingredient for obtaining a realistic description of quantum spectra. We account for the decay by using a Lindblad-type master equation with the decay rate $\gamma=1~\mu s^{-1}$ for each transmon. The evolution of the density matrix of the system is given by 
\begin{equation}
\label{eq:Lindblad}
\begin{split}
\frac{\partial \hat{\rho}}{\partial t} = -\frac{i}{\hbar}\left[\mathcal{\hat H}_\mathrm{RWA}, \hat{\rho}\right] + 
\sum\limits_{q=1}^{N} \frac{\gamma}{2}\, \left( 2\,\hat{a}_q \hat{\rho} \hat{a}_q^\dag - \hat{a}_q^\dag \hat{a}_q\hat{\rho} - \hat{\rho} \hat{a}_q^\dag \hat{a}_q \right)
\end{split}
\end{equation}
The drive amplitude used for these low-power simulations is $\Omega/2\pi = 1$~MHz. The spectrum presented in Fig.~2{\bf b} of the manuscript main text is simulated as the average occupancy of the first or the last transmon of the array for the steady state solution $\hat{\rho}_\mathrm{ss}$ of equation \eqref{eq:Lindblad}:
\begin{equation}
    \langle n \rangle = \operatorname{Tr}{\left[\hat{\rho}_\mathrm{ss} \hat{n}_s\right]}.
\end{equation}
Equation \eqref{eq:Lindblad} is a linear system of the first-order differential equations for the elements of the density matrix $\hat{\rho}$. The coefficient matrix of the system is the Liouvillian superoperator; the steady-state solutions are given by its nullspace.

For numerical computation of the steady-state density matrix, we truncate the Hilbert space of the 11-transmon system to the 78 eigenstates of the undriven Hamiltonian with the lowest energy containing zero, one or two photons. Accordingly, we neglect the contribution of the states with three or more excitations.

\section{Electric circuit model}

\begin{figure*}[hb]
	\centerline{\includegraphics[width=9cm]{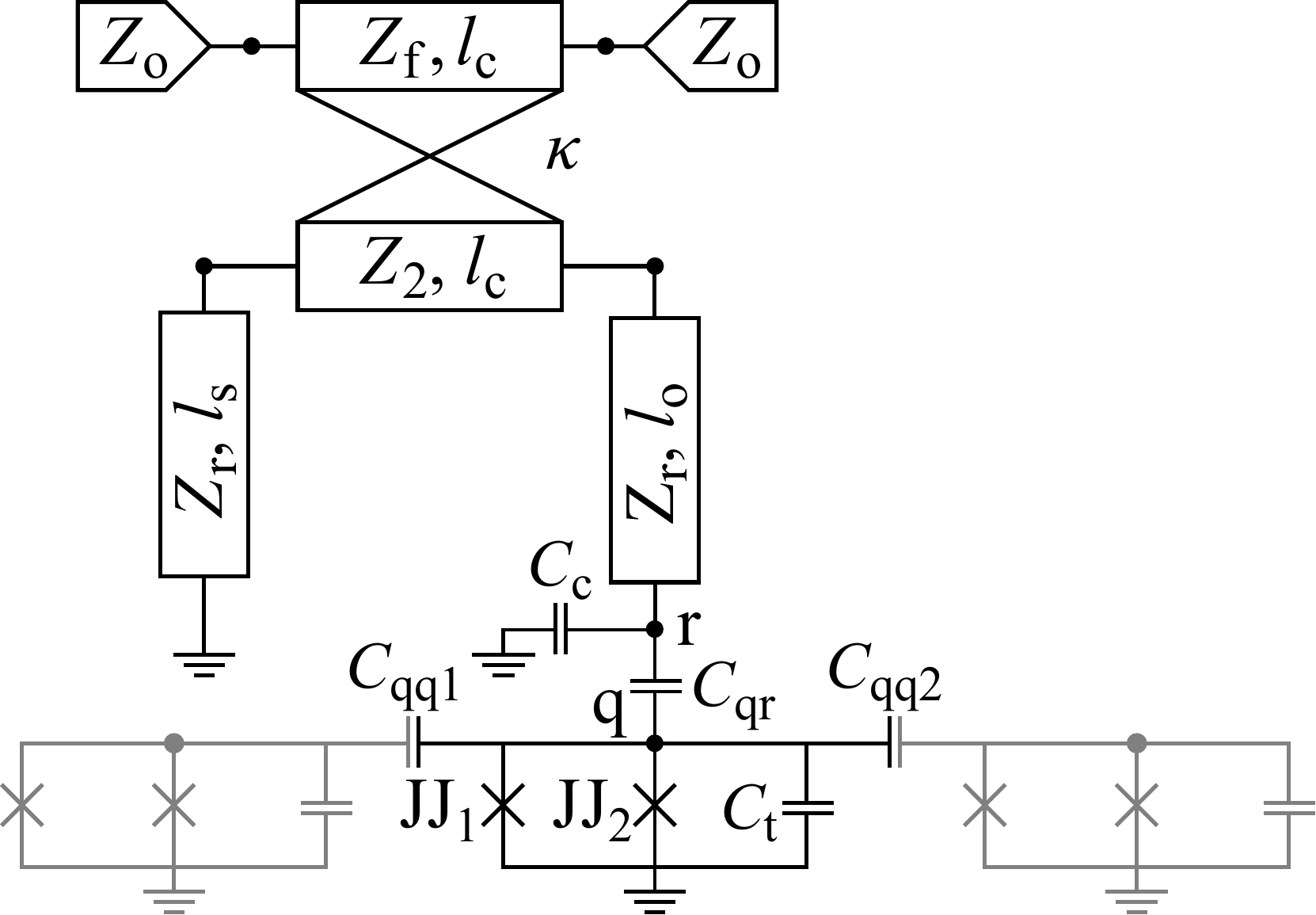}}
	\caption{\footnotesize {\bf Electric circuit model for the individual transmon and its readout resonator.} The transmon, the respective resonator and the readout feedline are shown in black. Neighboring transmons are shown in gray.}
	\label{im:electric-schematic}
\end{figure*}

The design values for the qubit frequency, qubit-qubit coupling, qubit-resonator coupling and resonator-feedline coupling have been calculated using the circuit model shown in Fig.~\ref{im:electric-schematic}. All couplings are purely of electrostatic nature, except for the resonator-feedline coupling which also has an inductive contribution. The lengths of the transmission line segments are $l_o=805\, \mu$m, $l_c=142\, \mu$m. The lengths of the shorted ends of the resonators $l_s$ are provided in Table~\ref{table:resonator-parameters}. The readout resonators' sections are realized as coplanar waveguides (CPWs) with conductor and gap widths of 10 $\mu$m and 6 $\mu$m, respectively. The feedline conductor width is 17~$\mu$m and the gap is 8~$\mu$m; in the coupler section, the distance between the feedline and resonator conductors is 8~$\mu$m. We take $\varepsilon=11.75$ as the dielectric constant of silicon and neglect the finite superconductor thickness (100 nm) for all calculations.

\begin{center}
\begin{table}
	\label{table:resonator-parameters}
	\begin{tabular}{| l | l | l | l | l | l | l | l | l | l | l | l |}
		\hline
		$q$ & $l_s, \mu m$ & $f_r$,~GHz & $Q_r$ & $g/\sqrt{f_q}, \sqrt{\mathrm{Hz}}$ & $f_r$,~GHz, experiment & $Q_r$, experiment \\ \hline
		
		1   & 3686 & 6.228  & 3862 & 852  & 6.260  & 3628  \\
		2   & 3645 & 6.275  & 2888 & 859  & 6.307  & 2224  \\
		3   & 3604 & 6.321  & 2894 & 866  & 6.363  & 2209  \\
		4   & 3564 & 6.368  & 2660 & 873  & 6.414  & 1975  \\
		5   & 3525 & 6.418  & 3377 & 880  & 6.459  & 1833  \\
		6   & 3493 & 6.447  & 1820 & 886  & 6.499  & 1199  \\
		7   & 3448 & 6.513  & 2476 & 894  & 6.555  & 1357  \\
		8   & 3410 & 6.561  & 2166 & 901  & 6.607  & 1136  \\
		9   & 3373 & 6.613  & 2092 & 908  & 6.659  & 1216  \\
		10  & 3337 & 6.666  & 2283 & 915  & 6.711  & 1331  \\
		11  & 3301 & 6.713  & 1264 & 922  & 6.763  & 761  \\
		\hline
	\end{tabular}
	\caption{Simulated and measured parameters of the readout resonators coupled to the 11 transmon qubits.}
\end{table}
\end{center}

We have extracted the mutual capacitances between the circuit elements using Ansys Maxwell software. Neglecting smaller capacitances of next-to-nearest couplings and other parasitic capacitances, we obtain the following values for the circuit model (the designations are shown in Fig.~\ref{im:electric-schematic}): $C_{\mathrm{c}} = 61.2$~fF, $C_{\mathrm{qr}} = 5.2$~fF, $C_{\mathrm{t}} = 119.1$~fF, $C_\mathrm{qq1} = 3.8$~fF, and $C_{\mathrm{qq2}} = 1.3$~fF.
The effective phase velocity for a thin film on a silicon-vacuum interface is 
$c_l = \frac{1}{\sqrt{\varepsilon_0\mu_0\left(\varepsilon+1\right)/2}}$, i.e. 2.52 times smaller than the speed of light in vacuum. The characteristic impedance of the uncoupled resonator segments obtained from conformal mapping is $Z_r=50.3\, \Omega$.
Small extra capacitance due to the qubit-resonator coupling leads to the increase of the effective length $l_o^{(e)}$ of the open-ended section of the resonator:
$$l_o^{(e)} = l_o + c_l Z_r \left(C_\mathrm{c}+C_{\mathrm{qr}}\right)$$
The qubit-resonator coupling constant is given by
$$g=\frac{1}{2}\sqrt{f_qf_r}
\frac{C_{\mathrm{qr}}}{\sqrt{C_{\mathrm{q}} C_{\mathrm{r}}}},$$
where the effective capacitance of the fundamental mode of the resonator is 
$$C_\mathrm{r} = \frac{l_s+l_c+l_o^{(e)}}{2Z_rc_l}. $$

We have used full-wave RF simulation with Ansys HFSS to obtain the resonator frequencies and external quality factors, as this yields significantly more accurate results than the electric circuit model.

For small coupling capacitances, the value of the anharmonicity can be approximated by
$$
\delta = -\frac{e^2}{2\left( C_\mathrm{t} + C_\mathrm{qr} + C_\mathrm{qq1} + C_\mathrm{qq2} \right)},
$$
The alternating qubit-qubit couplings are determined by the dimensionless coupling coefficients $$t_i = \frac{C_{\mathrm{qq}i}}{2\left( C_\mathrm{t} + C_\mathrm{qr} + C_\mathrm{qq1} + C_\mathrm{qq2} \right)}.$$








\section{Frequency control calibration}

\begin{figure*}[h!]
	\centerline{\includegraphics[width=0.9\linewidth]{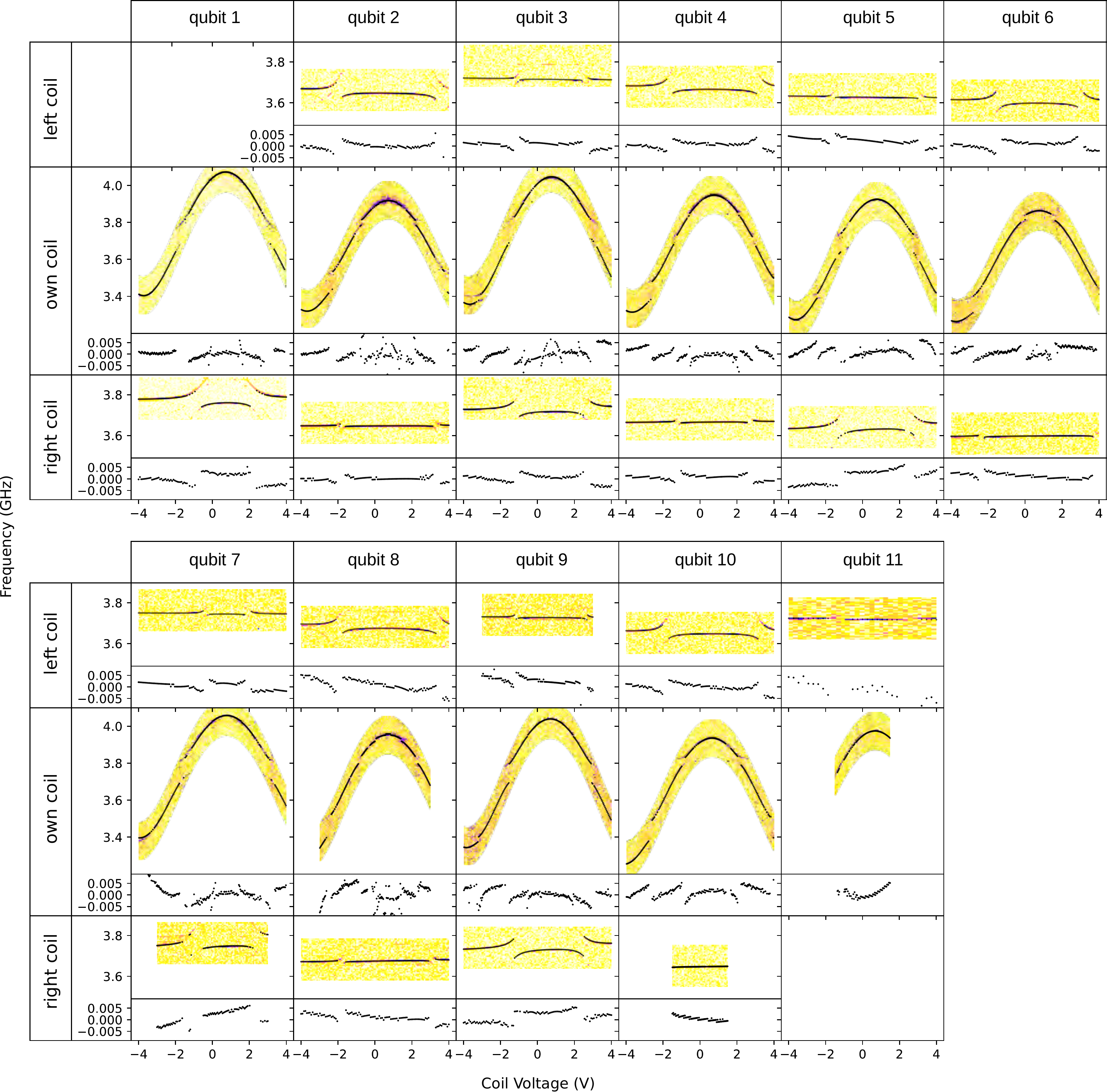}}
	\caption{\footnotesize {\bf Two-tone spectra used for calibration and fitting of the residuals.} Each column contains the result of two-tone spectroscopy recorded from a different resonator. Scans of the coil voltage of the transmon's own coil and its left and right neighbor's coils are provided in the different rows. Two-tone response is shown in color in arbitrary units. Fitted frequency values from the model are shown in black dots. Residuals are shown in panels under the corresponding spectra.}
	\label{im:two-tone-spectroscopy}
\end{figure*}

To extract the critical currents of the Josephson junctions, the mutual inductances between the flux lines, the respective transmons and the nearest-neighbor transmons as well as the flux biases of the SQUIDs (probably caused by stray magnetic fields frozen into the chip during the cooldown process) we perform two-tone spectroscopy on each of the resonators while scanning the flux on the corresponding line. The measurement is performed first at high excitation and readout power to approximately find the range of frequencies where the transmon frequency lies. After this preliminary measurement, we repeat the measurements with lower readout and excitation power with an adaptive frequency range. Additionally, we investigate the crosstalk between transmons and flux lines of the nearest neighbors' transmons by measuring corresponding two-tone spectra. The results of these measurements are presented in Fig.~\ref{im:two-tone-spectroscopy}.

For each external flux configuration, the maximum two-tone response frequency is taken as the transmon mode frequency corresponding to this configuration. Outlier frequencies are dropped, after which the transmon frequency dependence on field configuration is fitted by a coupled linear oscillator model. Eigenfrequencies of all transmons and resonators are determined as eigenvalues of a $22\times 22$ matrix which has the bare transmon and resonator frequencies as the diagonal elements and coupling strengths as off-diagonal elements. The values of bare resonator frequencies are found from the transmission minimum for the set of preliminary measurements. For the dependence of bare transmon frequency on SQUID flux we use the formula
\begin{equation}
	\label{eq:bare-fq}
	f_q(\Phi_q) = \sqrt{8E_C\sqrt{(E_{qJ1}+E_{qJ2})^2\cos^2\varphi_q+(E_{qJ1}-E_{qJ2})^2\sin^2\varphi_q}}-E_C,
\end{equation}
where $E_{qJ1}$ and $E_{qJ2}$ are the Josephson energies of the DC SQUIDs' junctions, $E_C$ is the transmon charging energy, $\varphi_q = 2\pi \Phi_q/\Phi_0$ is the dimensionless magnetic flux through the $q$-th SQUID.
We account for the nearest-neighbour flux line crosstalk by assuming linear dependence of the SQUID fluxes on DC voltage applied to the flux lines in series with a resistor with a tridiagonal inductance matrix $L_{ql}$:
\begin{equation}\label{eq:inductance-matrix}
	\Phi_q = \sum\limits_l L_{ql}V_l+\Phi^{(0)}_q,
\end{equation}
where $\Phi^{(0)}_q$ are the frozen-in fluxes in the SQUIDs due to the residual stray magnetic fields. 


The coupling coefficients between transmons and resonators are given by
\begin{equation}
	g_q(\Phi_q) = g\sqrt{f_q(\Phi_q)}.
\end{equation}

Finally, transmon-transmon couplings are approximated by the relation
\begin{equation}
    J_{q,(q+1)}=\sqrt{f_q f_{(q+1)}}\times
    \begin{cases}
        t_1, \text{if $q$ is odd,} \\
        t_2, \text{if $q$ is even}.
    \end{cases}
\end{equation}

The matrix constructed in this way has the form
\begin{equation}
    \begin{pmatrix}
        f_{R1}  &    0       & \dots & 0       & g_1        & 0         & \dots   & 0 \\
        0       &    f_{R2}  & \dots & 0       & 0        & g_2         & \dots   & 0 \\
        \vdots  &   \vdots   & \ddots& \vdots  & \vdots   & \vdots    & \ddots  & \vdots \\
        0       &    0       & \dots & f_{R11} & 0        & 0         & \dots   & g_{11} \\
        g_1       &    0       & \dots & 0       & f_{1}    & J_{1,2}    & \dots   & 0 \\
        0       &    g_2       & \dots & 0       & J_{1,2}   & f_{2}     & \dots   & 0 \\
        \vdots  &   \vdots   & \ddots& \vdots  & \vdots   & \vdots    & \ddots  & \vdots \\
        0       &    0       & \dots & g_{11}       & 0        & 0         & \dots   & f_{11} \\
    \end{pmatrix}
\end{equation}
The effective frequency of a specific transmon is chosen as the frequency of the mode with the largest participation in this transmon.

Fitting is performed with a least-squares cost function. The optimal fit parameters are shown in Table~\ref{table:fitresults}. The root mean square error of the fit is $11$~MHz. However, if outlier points with over 80 MHz discrepancy between numerical model and experiment are removed from the error calculation, the mean square error becomes 4 MHz.

To set the bare frequencies of all transmons to the desired values, we find the corresponding SQUID flux values using Eq.~\eqref{eq:bare-fq} and then solve the linear system of equations \eqref{eq:inductance-matrix}.

The finite resolution of the flux control DAC is equal to $\Delta V=2$~mV, which corresponds to an inherent average error of $\Delta f=0.2$~MHz. 

\begin{center}
\begin{table}
	\begin{tabular}[h!]{| l | l | l | l | l | l | l | l | l | l | l | l | l | l | l |}
		\hline
		$q$ & $f_r$,~GHz & $L_{q,q-1}, V/\Phi_0$ & $L_{q,q}, V/\Phi_0$ & $L_{q,q+1}, V/\Phi_0$ & $\Phi_q^{(0)}, \Phi_0$ & $E_{qJ1}$, GHz & $E_{qJ2}$, GHz                           \\ \hline
		1 & 6.260  &        & -0.1121 & -0.0005 & 0.1672 & 11.816 & 1.944   \\
		2 & 6.307  & 0.0008 & -0.1128 & -0.0001 & 0.1127 & 11.168 & 1.673   \\
		3 & 6.362  & 0.0006 & -0.1119 & -0.0006 & 0.1305 & 11.815 & 1.911   \\
		4 & 6.414  & 0.0004 & -0.1120 & -0.0003 & 0.0936 & 11.263 & 1.769   \\
		5 & 6.459  & 0.0006 & -0.1129 & -0.0017 & 0.1193 & 10.981 & 1.811   \\
		6 & 6.498  & 0.0003 & -0.1103 & -0.0003 & 0.1103 & 10.909 & 1.608   \\
		7 & 6.555  & 0.0003 & -0.1053 & -0.0022 & 0.1237 & 11.666 & 1.989   \\
		8 & 6.606  & 0.0007 & -0.1057 & -0.0005 & 0.1235 & 10.903 & 2.189   \\
		9 & 6.658  & 0.0003 & -0.1097 & -0.0017 & 0.1060 & 11.563 & 2.001   \\
		10 & 6.711 & 0.0001 & -0.1047 & -0.0005 & 0.0640 & 11.045 & 1.941   \\
		11 & 6.763 & 0.0005 & -0.1074 &         & 0.0756 & 11.265 & 1.946   \\
		\hline
	\end{tabular}
	\label{table:fitresults}
	\caption{Optimal fit parameters for the calibration model.}
\end{table}
\end{center}

\section{Evolution of the doublon edge state with the array length}

As we point out in the manuscript main text, in this work we observe the signatures of interaction-induced doublon edge states rather than the doublon edge state itself. The intrinsic reason for that is the insufficient length of the array. In this section, we examine the requirements for the observation of the doublon edge state on the strong link edge of the array.

\begin{figure*}[h]
	\centerline{\includegraphics[width=0.9\linewidth]{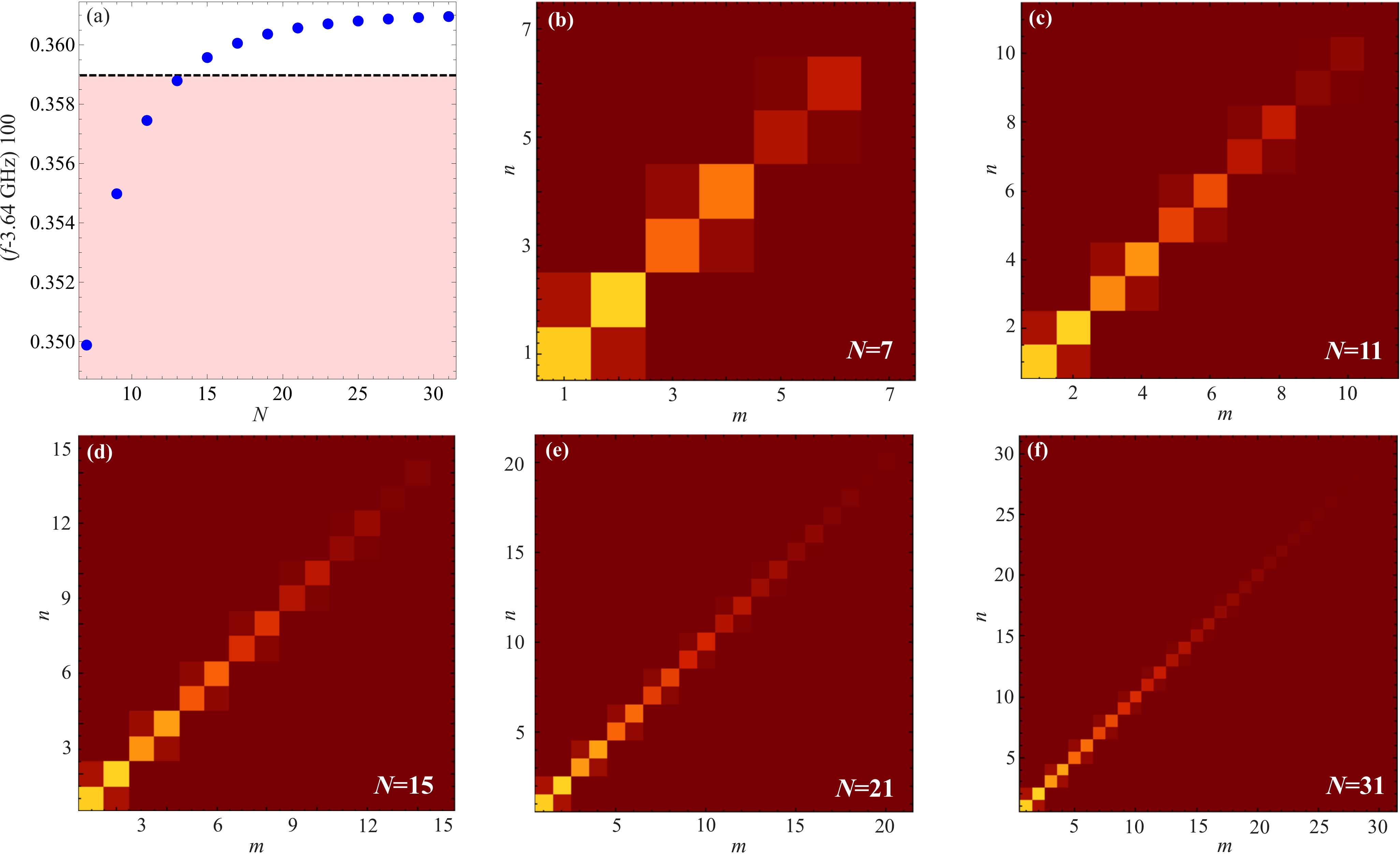}}
	\caption{\footnotesize {\bf Frequency and localization of the doublon edge state as a function of the array length.} (a) Frequency of the doublon edge state (blue dots) and the frequencies of bulk doublons (red shaded area) as a function of the array length. (b-e) Two-photon probability distributions for the doublon edge state calculated for the arrays with $N=7, 11, 15, 21$ and $31$ qubits.}
	\label{im:doublon-vs-N}
\end{figure*}

Diagonalizing the Bose-Hubbard Hamiltonian, we find out that for the lengths of the array $N<15$ the mode corresponding to the edge state appears in the bulk doublon band and hence its localization is elusive [Fig.~\ref{im:doublon-vs-N}(a)]. However, once $N>15$, the edge state appears to be outside of the bulk doublon band, while the respective two-photon probability distribution clearly shows the signatures of exponential localization [Fig.~\ref{im:doublon-vs-N}(b-f)].

\bibliographystyle{naturemag}
\bibliography{TopologicalLib1}